\documentclass[preprint,aps,amsmath,nofootinbib,tightenlines,showpacs]{revtex4}
\usepackage{amsmath}
\usepackage{amsfonts}
\usepackage{amssymb}
\usepackage{latexsym}
\usepackage{graphicx}
\usepackage{bm}
\usepackage{epsfig}
\usepackage[english]{babel}

\def\diag{\mbox{diag}}
\def\arcsinh{\mbox{arcsinh}}
\def\const{\mbox{const}}
\def\mass{\mbox{mass}}

\begin{document}


\title{Static Solutions with Spherical Symmetry in $f(T)$ Theories}
\author{Tower Wang\footnote{Electronic address: twang@phy.ecnu.edu.cn}}
\affiliation{Department of Physics, East China Normal University,\\
Shanghai 200241, China\\ \vspace{0.2cm}}
\date{\today\\ \vspace{1cm}}
\begin{abstract}
The spherically symmetric static solutions are searched for in some
$f(T)$ models of gravity theory with a Maxwell term. To do this, we
demonstrate that reconstructing the Lagrangian of $f(T)$ theories is
sensitive to the choice of frame, and then we introduce a particular
frame based on the conformally Cartesian coordinates. In this
particular frame, the existence conditions of various solutions are
presented. Our results imply that only a limited class of $f(T)$
models can be solved in this frame. For more general models, the
search for spherically symmetric static solutions is still an open
and challenging problem, hopefully solvable in other frames.
\end{abstract}


\maketitle



\section{Introduction and Preliminaries}\label{sect-intro}
As a great triumph of science in the last century, Einstein's theory
of general relativity with a cosmological constant has been
established by many observations and experiments hitherto. In spite
of this fact, people have been enthusiastically trying to modify or
extend this theory. Cosmologically the motivation to do it is
running our Universe with less dark matter
\cite{Milgrom:1983ca,Bekenstein:2004ne} or without the cosmological
constant \cite{Carroll:2003wy,Nojiri:2003ft,Nojiri:2006ri}, or
driving inflation without a scalar field
\cite{Starobinsky:1980te,DeFelice:2010aj} or with richer phenomena
\cite{Pi:2009an}.

A few years ago, $f(T)$ theories of gravity were proposed as an
alternative of the cosmological constant to explain the accelerated
expansion of the late Universe \cite{Bengochea:2008gz}. Similar to
the $f(R)$ theories, $f(T)$ theories deviate from Einstein gravity
by a function $f(T)$ in the Lagrangian, where $T$ is the so-called
torsion scalar. This class of theories has received a lot of
attention recently
\cite{Linder:2010py,Wu:2010mn,Myrzakulov:2010vz,Yerzhanov:2010vu,Wu:2010xk,Yang:2010hw,Tsyba:2010ji,Chen:2010va,Bengochea:2010sg,Wu:2010av,Bamba:2010iw,Myrzakulov:2010tc,Karami:2010bu,Karami:2010xy,Li:2010cg,Yang:2010ji,Dent:2010bp,Zheng:2010am,Bengochea:2010it,Bamba:2010wb,Sotiriou:2010mv},
but earlier examples could be traced to
\cite{Ferraro:2006jd,Ferraro:2008ey}. In order to better study
$f(T)$ gravity theories, we will seek for exact solutions of the
field equations, akin to the work
\cite{Briscese:2007cd,delaCruzDombriz:2009et,delaCruzDombriz:2010vc,Sebastiani:2010kv,Moon:2011hq}
done in $f(R)$ models. Especially we are interested in spherically
symmetric solutions. The widely studied
Friedmann-Lema\^{i}tre-Robertson-Walker (FLRW) metric is such a
solution obviously, but we will pay attention to static solutions
instead.

In this paper, the notation of indices is as follows: Greek indices
$\mu,\nu,...$ and Latin indices from the beginning of the alphabet
$a,b,...$ run over $0,1,2,3$, while Latin indices from the middle of
the alphabet $i,j,...$ run from 1 to 3. In our notations, the
Kronecker delta is defined with the same normalization irrespective
of the position of indices, \emph{e.g.}
$\delta_{00}=\delta^{00}=\delta^0_0=1$.

The dynamic variables in $f(T)$ theories are the tetrad (vierbein)
fields $\mathbf{e}_a$. In a coordinate basis, the components of dual
tetrad fields $e^a_{\mu}dx^{\mu}$ are related to the familiar metric
tensor $g_{\mu\nu}$ through
\begin{equation}
g_{\mu\nu}=\eta_{ab}e^a_{\mu}e^b_{\nu},
\end{equation}
where $\eta_{ab}=\diag(1,-1,-1,-1)$. Starting from the tetrad
fields, one may follow \cite{Bengochea:2008gz,Chen:2010va,Li:2010cg}
to construct the torsion tensor
\begin{equation}\label{T-tensor}
T^{\lambda}_{~\mu\nu}=e^{\lambda}_a(\partial_{\mu}e^a_{\nu}-\partial_{\nu}e^a_{\mu}),
\end{equation}
the contorsion tensor
\begin{equation}\label{K-tensor}
K^{\mu\nu}_{~~\lambda}=-\frac{1}{2}(T^{\mu\nu}_{~~\lambda}-T^{\nu\mu}_{~~\lambda}-T_{\lambda}^{~\mu\nu})
\end{equation}
and a useful tensor
\begin{equation}\label{S-tensor}
S_{\lambda}^{~\mu\nu}=\frac{1}{2}(K^{\mu\nu}_{~~\lambda}+\delta^{\mu}_{\lambda}T^{\rho\nu}_{~~\rho}-\delta^{\nu}_{\lambda}T^{\rho\mu}_{~~\rho}).
\end{equation}
Then in terms of the torsion scalar
\begin{equation}\label{T-scalar}
T=S_{\lambda}^{~\mu\nu}T^{\lambda}_{~\mu\nu}
\end{equation}
and the notation $e=\det(e^a_{\mu})=\sqrt{-g}$, the action of $f(T)$
gravity theories is given by
\begin{equation}\label{action}
I=\int d^4xe\left[\frac{1}{16\pi G}(T+f)+\mathcal{L}_m\right],
\end{equation}
which leads to the equations of motion
\begin{equation}\label{eom}
(1+f_{,T})[e^{-1}\partial_{\mu}(eS_{a}^{~\mu\nu})-e^{\lambda}_aT^{\rho}_{~\mu\lambda}S_{\rho}^{~\nu\mu}]+f_{,TT}S_{a}^{~\mu\nu}\partial_{\mu}T-\frac{1}{4}e^{\nu}_a(T+f)=4\pi Ge^{\rho}_a\mathcal{T}_{\rho}^{~\nu}
\end{equation}
with $S_{a}^{~\mu\nu}=e^{\rho}_aS_{\rho}^{~\mu\nu}$. If the function
$f(T)$ is replaced by a constant, the theory is equivalent to
Einstein's theory with a cosmological constant \cite{Li:2010cg}.

Throughout this paper, the Maxwell term
\begin{equation}\label{Maxwell}
\mathcal{L}_m=-\frac{1}{4}F_{\mu\nu}F^{\mu\nu}
\end{equation}
will be considered. It gives the energy-momentum tensor
\begin{equation}
\mathcal{T}_{\rho}^{~\nu}=F_{\rho\lambda}F^{\nu\lambda}-\frac{1}{4}\delta_{\rho}^{\nu}F_{\lambda\mu}F^{\lambda\mu}.
\end{equation}
We will need the Ricci scalar $R=g^{\mu\nu}R_{\mu\nu}$, taking the
convention of Ricci tensor
\begin{equation}\label{R-tensor}
R_{\mu\nu}=\partial_{\lambda}\Gamma^{\lambda}_{\mu\nu}-\partial_{\nu}\Gamma^{\lambda}_{\mu\lambda}+\Gamma^{\lambda}_{\lambda\rho}\Gamma^{\rho}_{\mu\nu}-\Gamma^{\lambda}_{\nu\rho}\Gamma^{\rho}_{\mu\lambda}
\end{equation}
with the Levi-Civita connection
\begin{equation}\label{Levi-Civita}
\Gamma^{\lambda}_{\mu\nu}=\frac{1}{2}g^{\lambda\rho}(\partial_{\mu}g_{\rho\nu}+\partial_{\nu}g_{\mu\rho}-\partial_{\rho}g_{\mu\nu}).
\end{equation}

In the rest of this paper, after choosing the frame in section
\ref{sect-frame}, we will search for the spherically symmetric
static solutions in different situations in sections
\ref{sect-hunt}, \ref{sect-regular}, \ref{sect-horizon}. Then we
will recap the solutions and their constraints on Lagrangian in
section \ref{sect-lag}. In section \ref{sect-concl} some open issues
will be mentioned.

\section{The Choice of Frame}\label{sect-frame}
The first difficulty we encountered is the choice of frame. As
dynamic variables in $f(T)$ theories, tetrad fields are sensitive to
the frame. More importantly, the torsion scalar is also
frame-sensitive, very much unlike the Ricci scalar. This trouble is
attributed to the lack of local Lorentz invariance in $f(T)$
theories \cite{Li:2010cg}. Consequently, even for the same metric
and the same coordinate basis, different frames result in different
forms of equations of motion.

To see this point clearly, we take the FLRW metric for example.
Corresponding to the coordinates
\begin{equation}\label{FLRWc}
ds^2=dt^2-a^2(t)\delta_{ij}dx^idx^j,
\end{equation}
the following form of tetrads are widely used in the literature
\cite{Bengochea:2008gz,Linder:2010py},
\begin{equation}\label{tetrc1}
e^0_{\mu}dx^{\mu}=\delta^0_{\mu}dx^{\mu},~~~~e^i_{\mu}dx^{\mu}=a\delta^i_{\mu}dx^{\mu}.
\end{equation}
It give the torsion scalar $T=-6H^2$ and the Ricci scalar
$R=-6(2H^2+\dot{H})$. Here the Hubble parameter $H=\dot{a}/a$, and
the dot overhead implies the derivative with respect to $t$. Then
equations of motion \eqref{eom} lead to the generalized Friedmann
equations
\begin{eqnarray}\label{Friedmannc}
\nonumber f+6H^2+12H^2f_{,T}&=&16\pi G\rho,\\
f+6H^2+4\dot{H}+4(3H^2+\dot{H})f_{,T}-48H^2\dot{H}f_{,TT}&=&-16\pi Gp.
\end{eqnarray}
Expression \eqref{tetrc1} is an obvious form of tetrads yielding
metric \eqref{FLRWc}. But it is not the unique choice. For instance,
we can rotate it to a different frame, and write down the following
form of tetrads:
\begin{eqnarray}\label{tetrc2}
\nonumber e^0_{\mu}dx^{\mu}&=&dt,\\
\nonumber e^1_{\mu}dx^{\mu}&=&\frac{a}{r}(xdx+ydy+zdz),\\
\nonumber e^2_{\mu}dx^{\mu}&=&\frac{az}{r\sqrt{x^2+y^2}}(xdx+ydy)-\frac{a\sqrt{x^2+y^2}}{r}dz,\\
e^3_{\mu}dx^{\mu}&=&-\frac{a}{\sqrt{x^2+y^2}}(ydx-xdy),
\end{eqnarray}
in which $r=(x^2+y^2+z^2)^{1/2}$. It is easy to check that the new
form of tetrads can also produce metric \eqref{FLRWc} and the local
rotation matrix
\begin{equation}\label{rotation}
\begin{pmatrix}
\frac{x}{r}&\frac{y}{r}&\frac{z}{r}\\
\frac{xz}{r\sqrt{x^2+y^2}}&\frac{yz}{r\sqrt{x^2+y^2}}&-\frac{\sqrt{x^2+y^2}}{r}\\
-\frac{y}{\sqrt{x^2+y^2}}&\frac{x}{\sqrt{x^2+y^2}}&0\\
\end{pmatrix}
\end{equation}
transforms \eqref{tetrc1} to \eqref{tetrc2}. Keep in mind that the
rotation group is a subgroup of the proper Lorentz transformation,
so frames \eqref{tetrc1} and \eqref{tetrc2} are related by a local
Lorentz transformation. In accordance with tetrads \eqref{tetrc2},
the torsion scalar becomes $T=2a^{-2}r^{-2}-6H^2$ but the Ricci
scalar remains $R=-6(2H^2+\dot{H})$. Now equations of motion
\eqref{eom} reduce to
\begin{eqnarray}\label{Friedmanns}
\nonumber f+6H^2+12H^2f_{,T}-\frac{2}{a^2r^2}f_{,T}&=&16\pi G\rho,\\
\nonumber f_{,TT}&=&0,\\
f+6H^2+4\dot{H}+4(3H^2+\dot{H})f_{,T}-\frac{2}{a^2r^2}f_{,T}&=&-16\pi Gp.
\end{eqnarray}
Comparing \eqref{Friedmannc} and \eqref{Friedmanns}, we can see the
equations of motion, as well as the reconstruction of $f(T)$, is
quite sensitive to the choice of frame. Especially, the second
equation of \eqref{Friedmanns} requires $f=\lambda T-2\Lambda$.
After rescaling the Newtonian constant, the resulted theory is
simply equivalent to Einstein's theory with a cosmological constant.
But the commonly studied equations \eqref{Friedmannc}, which are
derived in frame \eqref{tetrc1}, allow much more general forms of
$f(T)$ to survive. This property does not rely on the choice of
coordinate system\footnote{The author is indebted to Rong-Xin Miao
and the referee for their valuable comments on this point.}. Through
the coordinate transformation $x=r\sin\theta\cos\phi$,
$y=r\sin\theta\sin\phi$, $z=r\cos\theta$, expression \eqref{tetrc2}
is traded to
\begin{equation}\label{tetrs}
e^0_{\mu}dx^{\mu}=dt,~~~~e^1_{\mu}dx^{\mu}=adr,~~~~e^2_{\mu}dx^{\mu}=ard\theta,~~~~e^3_{\mu}dx^{\mu}=ar\sin\theta d\phi
\end{equation}
in a spherical coordinate system
\begin{equation}\label{FLRWs}
ds^2=dt^2-a^2(t)(dr^2+r^2d\Omega^2)
\end{equation}
with $d\Omega^2=d\theta^2+\sin^2\theta d\phi^2$. That is to say,
\eqref{tetrc2} and \eqref{tetrs} describe the same frame in
different coordinate bases.

The above example sharpens the problem of what frame we should adopt
to look for new solutions. It is interesting to notice that the
spatial coordinates in metric \eqref{FLRWc} are Cartesian up to a
conformal factor. This coordinate system is relatively more
convenient for us to perform the Lorentz boost and the rotation of
spatial axes locally. Fortunately, there is a close cousin of
\eqref{FLRWc}, namely the generalized Gullstrand-Painlev\'e metric
\begin{equation}\label{GPmetric}
ds^2=dt^2-\delta_{ij}\left(\frac{\sqrt{\alpha}x^i}{r}dt+\frac{dx^i}{\beta}\right)\left(\frac{\sqrt{\alpha}x^j}{r}dt+\frac{dx^j}{\beta}\right),
\end{equation}
where $\alpha$ and $\beta$ are functions of the radial coordinate
$r=(\delta_{ij}x^ix^j)^{1/2}$ for static solutions. In this
coordinate system, every $x^{\mu}$ fully spans $(-\infty,+\infty)$,
and the spatial coordinates are conformally Cartesian
\cite{Damour:1990pi,Vines:2010ca}. Through a coordinate
transformation
\begin{equation}
t=\tilde{t}+\int\frac{\sqrt{\alpha}}{\beta(1-\alpha)}dr,
\end{equation}
metric \eqref{GPmetric} can be rewritten in the more familiar form
\begin{eqnarray}\label{Schmetric}
\nonumber ds^2&=&dt^2-\left(\sqrt{\alpha}dt+\frac{dr}{\beta}\right)^2-\frac{r^2}{\beta^2}d\Omega^2\\
&=&(1-\alpha)d\tilde{t}^2-\frac{1}{\beta^2(1-\alpha)}dr^2-\frac{r^2}{\beta^2}d\Omega^2
\end{eqnarray}
in spherical coordinates.

From the metric in Gullstrand-Painlev\'e coordinates, we directly
read out one possible form of the tetrad fields
\begin{equation}\label{tetrad}
e^0_{\mu}dx^{\mu}=dt,~~~~e^i_{\mu}dx^{\mu}=\frac{\sqrt{\alpha}x^i}{r}dt+\frac{1}{\beta}dx^i.
\end{equation}
This is the closest analogue of \eqref{tetrc1} for the solution we
are interested in. It is also the starting point of our
investigation. The impact of such a choice of frame is twofold. On
the one hand, it simplifies our calculation significantly, otherwise
the Lorentz transformation will induce six more variables and then
the equations of motion would be too cumbersome to tackle. On the
other hand, it restricts the validity of our solutions to a limited
class of $f(T)$ models, while the search for spherically symmetric
static solutions in more general models remains a challenging task.

\section{Hunting for Spherically Symmetric Static Solutions}\label{sect-hunt}
In the Maxwell term \eqref{Maxwell},
$F_{\mu\nu}=\partial_{\mu}A_{\nu}-\partial_{\nu}A_{\mu}$ for an
electromagnetic field. If the field is static and spherically
symmetric, we can make use of $U(1)$ gauge invariance to write it in
the form $A_{\mu}dx^{\mu}=\gamma(r)dt$. The electric charge of our
solution can be obtained by evaluating the integral
\begin{equation}\label{charge}
-\frac{1}{4\pi}\int\sqrt{-g}g^{tt}g^{rr}F_{tr}d\theta d\varphi=-\frac{r^2\gamma_{,r}}{\beta}
\end{equation}
in the limit $r\rightarrow\infty$.

Starting from tetrads of the form \eqref{tetrad}, we find the
torsion scalar
\begin{equation}\label{Ts}
T=-\frac{2}{r^2}[\alpha(\beta-r\beta_{,r})^2+r\beta\alpha_{,r}(\beta-r\beta_{,r})-r^2\beta_{,r}^2].
\end{equation}
The equations of motion are complicated, whose tedious form will not
be shown here. We succeeded in casting them in simple forms under
different situations. One situation is that with $\alpha=0$. This
will be studied in section \ref{sect-regular}. The other situation
is to be investigated in section \ref{sect-horizon}, where
$\alpha\neq0$.

\section{$\alpha=0$, $\beta\neq0$}\label{sect-regular}
For physical solutions, $\beta$ should be non-zero, but $\alpha=0$
is still allowed. In this situation, the torsion scalar simplifies
as $T=2\beta_{,r}^2$, and the equations of motion reduce to
\begin{eqnarray}
\label{eom1a}r\beta\left[\frac{\beta_{,r}}{r}(1+f_{,T})\right]_{,r}-\left(\frac{T}{2}-\frac{2\beta\beta_{,r}}{r}\right)(1+f_{,T})&=&0,\\
\label{eom1b}\left(2T-\frac{4\beta\beta_{,r}}{r}\right)(1+f_{,T})-(T+f)+8\pi G\beta^2\gamma_{,r}^2&=&0,\\
\label{eom1c}r\beta\left[\frac{\beta_{,r}}{r}(1+f_{,T})\right]_{,r}+8\pi G\beta^2\gamma_{,r}^2&=&0,\\
\label{eom1d}\left(\frac{r^2\gamma_{,r}}{\beta}\right)_{,r}&=&0.
\end{eqnarray}
Equation \eqref{eom1a} can be integrated, resulting in the relation
\begin{equation}\label{eom1aa}
\frac{r\beta_{,r}}{\beta}(1+f_{,T})=\frac{8\pi GQ^2}{r_{\infty}^2}
\end{equation}
Here $Q$ and $r_{\infty}$ are constants, whose physical significance
will be clear later. We will assume
$r_{\infty}>0$ without loss of generality. Inserting this relation
into \eqref{eom1c}, we get
\begin{equation}\label{eom1cc}
\frac{Q^2}{r_{\infty}^2}\left(\frac{r\beta_{,r}}{\beta}-2\right)+r^2\gamma_{,r}^2=0
\end{equation}
because $\beta\neq0$. This equation can be combined with
\eqref{eom1d} to give a second-order ordinary differential equation
of $\gamma$,
\begin{equation}\label{eom1dd}
\frac{Q^2}{r_{\infty}^2}\gamma_{,rr}+r\gamma_{,r}^3=0,
\end{equation}
which leads to
\begin{equation}\label{eom1ddd}
\gamma_{,r}=-\frac{Q}{r_{\infty}\sqrt{r^2+r_0^2}}
\end{equation}
with a nonnegative constant $r_0$. We have chosen a proper signature
to ensure that $Q$ is the electric charge given by \eqref{charge}.
Now the $\gamma_{,r}^2$ term can be eliminated in equation
\eqref{eom1cc} and the equation becomes
\begin{equation}\label{eom1aaa}
\frac{Q^2}{r_{\infty}^2}\left(\frac{r\beta_{,r}}{\beta}-\frac{r^2+2r_0^2}{r^2+r_0^2}\right)=0.
\end{equation}
When $Q/r_{\infty}\neq0$, this is a first-order equation of $\beta$.
We will continue our discussion in two situations: $Q/r_{\infty}=0$
in subsection \ref{subsect-MinkRRS} and $Q/r_{\infty}\neq0$ in
subsection \ref{subsect-RRS}.

\subsection{$Q/r_{\infty}=0$}\label{subsect-MinkRRS}
In the situation $Q/r_{\infty}=0$, equation \eqref{eom1ddd}
indicates $\gamma=0$ up to a constant eliminable by $U(1)$ gauge
transformation. At the same time, the right hand side of equation
\eqref{eom1aa} vanishes. To this equation there are two solutions.

One solution is a constant $\beta$, and we fix it to be $\beta=1$ by
rescaling the spatial coordinates $x^i$. This solution, summarized
as
\begin{eqnarray}\label{sol1a}
\nonumber \alpha&=&0,\\
\nonumber \beta&=&1,\\
\gamma&=&0,
\end{eqnarray}
describes a Minkowski spacetime. One may check the torsion scalar
$T=0$ and the Ricci scalar $R=0$. It is easy to also check that the
constant solution satisfies \eqref{eom1a}, \eqref{eom1c} and
\eqref{eom1d} automatically. But the solution is incompatible with
equation \eqref{eom1b} unless $f(0)=0$. That does not mean the
gravity theory must recede to the Einstein's theory to guarantee the
existence of global Minkowski solution. In fact, it provides a
sufficient but not necessary condition: the global Minkowski
solution exists if function $f(T)$ vanishes at point $T=0$
numerically. If this condition is violated, the Minkowski solution
might still exist in frames other than \eqref{tetrad}.

The other solution to \eqref{eom1aa} is
\begin{equation}\label{fT}
1+f_{,T}=0.
\end{equation}
Then equation \eqref{eom1b} tells us
\begin{equation}\label{f}
T+f=0.
\end{equation}
If $T$ is not a constant, one should solve equations \eqref{fT} and
\eqref{f} analytically as differential equations. The only
consistent solution to them is $f=-T$ analytically. However, this
solution is unphysical and should be ruled out, otherwise the theory
of gravity is ill-defined by a null Lagrangian. If $T$ is a
constant, it will be enough to require these equations to hold
numerically, not analytically. In other words, we do not have to
solve them as differential equations. Remembering that
$T=2\beta_{,r}^2$, a constant $T$ can be achieved by
\begin{eqnarray}\label{sol1b}
\nonumber \alpha&=&0,\\
\nonumber \beta&=&\frac{r}{r_{\infty}}+k,\\
\gamma&=&0.
\end{eqnarray}
Here the constant $k=0$ or $\pm1$ after rescaling $x^i$. From
\eqref{sol1b} we work out the torsion scalar $T=2/r_{\infty}^2$ and
the Ricci scalar
\begin{equation}\label{Ricci1b}
R=-\frac{2}{r_{\infty}^2}-\frac{8k}{r_{\infty}r}.
\end{equation}
Clearly there is a curvature singularity at $r=0$ unless
$r_{\infty}$ is infinite or $k=0$. So let us consider different
choices of parameters.
\begin{enumerate}
\item $r_{\infty}$ is infinite and $k=0$. This solution gives $\beta=0$ and
an ill-behaved metric.
\item $r_{\infty}$ is infinite and $k=\pm1$. We identify this solution as the
Minkowski solution \eqref{sol1a}, but emphasize that equation
\eqref{fT} is not a necessary condition for the existence of
Minkowski solution, as shown at the beginning of this subsection.
\item $r_{\infty}$ is finite and $k=0$. In terms of a new
coordinate $z=r_{\infty}\ln(r/r_{\infty})$, the metric from this
solution takes the form
\begin{equation}\label{metric1a}
ds^2=d\tilde{t}^2-dz^2-r_{\infty}^2d\Omega^2.
\end{equation}
This metric describes a spacetime of topology
$\mathbf{R}^1\times\mathbf{R}^1\times\mathbf{S}^2$, where the
$\mathbf{S}^2$ sphere is of constant radius $r_{\infty}$.
\item $r_{\infty}$ is finite and $k=\pm1$. It is easy to
see the curvature singularity at $r=0$ is a naked singularity in
these solution.
\end{enumerate}

\subsection{$Q/r_{\infty}\neq0$}\label{subsect-RRS}
In situations with $Q/r_{\infty}\neq0$, equation \eqref{eom1aaa} is
solved by
\begin{equation}\label{eom1ccc}
\beta=\frac{r^2}{r_{\infty}\sqrt{r^2+r_0^2}},
\end{equation}
which immediately gives
\begin{equation}\label{T1ccc}
T=\frac{2r^2(r^2+2r_0^2)^2}{r_{\infty}^2(r^2+r_0^2)^3}.
\end{equation}
One may use \eqref{charge}, \eqref{eom1ddd} and \eqref{eom1ccc} to
prove that indeed $Q$ is the electric charge.

Substituting equations \eqref{eom1ddd}, \eqref{eom1aaa},
\eqref{eom1ccc} and \eqref{T1ccc} into \eqref{eom1b} and
\eqref{eom1aa}, we obtain
\begin{eqnarray}\label{eom1bb}
\nonumber T+f&=&\frac{8\pi GQ^2r^2(r^2+4r_0^2)}{r_{\infty}^4(r^2+r_0^2)^2},\\
1+f_{,T}&=&\frac{8\pi GQ^2(r^2+r_0^2)}{r_{\infty}^2(r^2+2r_0^2)}.
\end{eqnarray}
Making use of \eqref{T1ccc}, one may check the consistency condition
$(T+f)_{,r}=(1+f_{,T})T_{,r}$ by straightforward calculations.

\begin{figure}
\begin{center}
\includegraphics[width=0.45\textwidth]{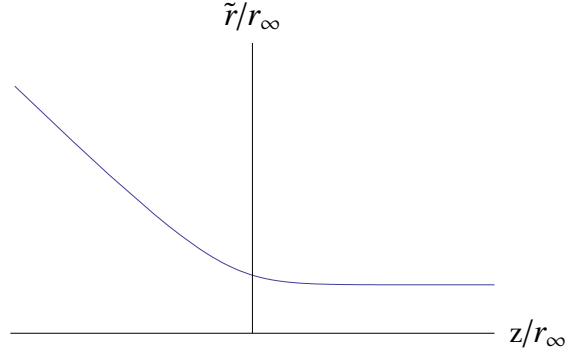}\\
\end{center}
\caption{The radius $\tilde{r}$  of $\mathbf{S}^2$ \eqref{radius}
decreases monotonically as coordinate $z$ \eqref{z}
increases.}\label{fig-radius}
\end{figure}

In order to work out more details, we should consider two
possibilities.
\begin{enumerate}
\item $r_0=0$. For such a choice, we get the solution
\begin{eqnarray}\label{sol1c}
\nonumber \alpha&=&0,\\
\nonumber \beta&=&\frac{r}{r_{\infty}},\\
\gamma&=&-\frac{Q}{r_{\infty}}\ln\left(\frac{r}{r_{\infty}}\right).
\end{eqnarray}
In the notation $z=r_{\infty}\ln(r/r_{\infty})$, it gives metric
\eqref{metric1a} and $\gamma=-Qz/r_{\infty}^2$. Furthermore, equations
\eqref{eom1bb} become
\begin{eqnarray}\label{eom1bbb}
\nonumber T+f&=&\frac{8\pi GQ^2}{r_{\infty}^4},\\
1+f_{,T}&=&\frac{8\pi GQ^2}{r_{\infty}^2},
\end{eqnarray}
which should hold numerically for $T=2/r_{\infty}^2$. Extrapolated
to the limit $Q=0$, this solution reproduces the regular
$\mathbf{R}^1\times\mathbf{R}^1\times\mathbf{S}^2$ solution
presented in the previous subsection.
\item $r_0>0$. Then as a cubic equation of the variable
$r^2/r_0^2$, equation \eqref{T1ccc} has one real root
$r^2/r_0^2=u(T)$ and thus
\begin{eqnarray}\label{eom1bbbb}
\nonumber T+f&=&\frac{8\pi GQ^2u(u+4)}{r_{\infty}^4(u+1)^2},\\
1+f_{,T}&=&\frac{8\pi GQ^2(u+1)}{r_{\infty}^2(u+2)}.
\end{eqnarray}
The metric is decided by the solution
\begin{eqnarray}\label{sol1d}
\nonumber \alpha&=&0,\\
\nonumber \beta&=&\frac{r^2}{r_{\infty}\sqrt{r^2+r_0^2}},\\
\gamma&=&\frac{Q}{r_{\infty}}\arcsinh\left(\frac{r_{\infty}}{r_0}\right)-\frac{Q}{r_{\infty}}\arcsinh\left(\frac{r}{r_0}\right).
\end{eqnarray}
In terms of coordinate
\begin{equation}\label{z}
z=r_{\infty}\left[\arcsinh\left(\frac{r}{r_0}\right)-\arcsinh\left(\frac{r_{\infty}}{r_0}\right)+\sqrt{1+\frac{r_0^2}{r_{\infty}^2}}-\sqrt{1+\frac{r_0^2}{r^2}}\right],
\end{equation}
we write it as
\begin{equation}\label{metric1b}
ds^2=d\tilde{t}^2-dz^2-\tilde{r}^2d\Omega^2,
\end{equation}
which describes a spacetime of topology
$\mathbf{R}^1\times\mathbf{R}^1\times\mathbf{S}^2$. The radius of
$\mathbf{S}^2$ sphere is
\begin{equation}\label{radius}
\tilde{r}=r_{\infty}\sqrt{1+\frac{r_0^2}{r^2}}.
\end{equation}
In the region $r\in(0,\infty)$, both $z$ and $\tilde{r}$ are strict
monotonic functions of $r$, so the radius $\tilde{r}$ shrinks
monotonically as $z$ varies from $-\infty$ to $+\infty$, as
illustrated in figure \ref{fig-radius}. Extrapolating this solution
to the limit $r_0=0$, we recover equations \eqref{metric1a},
\eqref{sol1c}, \eqref{eom1bbb} as well as $T=2/r_{\infty}^2$ and
$z=r_{\infty}\ln(r/r_{\infty})$. Note the extrapolation from
\eqref{eom1bbbb} to \eqref{eom1bbb} is a little tricky: the special
metric \eqref{metric1a} exists as long as equations \eqref{eom1bbb}
hold numerically at the point $T=2/r_{\infty}^2$, but
\eqref{eom1bbbb} should stand analytically to guarantee the
existence of the general metric \eqref{metric1b}.
\end{enumerate}

\section{$\alpha\beta\neq0$}\label{sect-horizon}
When $\alpha\beta\neq0$, the equations of motion can be arranged in
the form
\begin{eqnarray}
\label{eom2a}(1+f_{,T})\left(\frac{r\beta_{,r}}{\beta}\right)_{,r}&=&0,\\
\label{eom2b}f_{,TT}T_{,r}&=&0,\\
\label{eom2c}\left(2T-\frac{4\beta\beta_{,r}}{r}\right)(1+f_{,T})-(T+f)+8\pi G\beta^2\gamma_{,r}^2&=&0,\\
\label{eom2d}\left[2\alpha\left(\frac{\beta}{r}-\beta_{,r}\right)^2+\beta_{,r}\left(\frac{4\beta}{r}-\alpha_{,r}\beta-2\beta_{,r}\right)-\alpha_{,rr}\beta^2\right](1+f_{,T})-16\pi G\beta^2\gamma_{,r}^2&=&0,\\
\label{eom2e}\left(\frac{r^2\gamma_{,r}}{\beta}\right)_{,r}&=&0.
\end{eqnarray}
The last equation is the non-vanishing component of Maxwell
equations.

One special case is $f_{,T}=-1$. Keep in mind that $\beta\neq0$,
then equation \eqref{eom2d} indicates $\gamma=0$ up to $U(1)$ gauge
transformation.

If $f_{,T}\neq-1$, equations\eqref{eom2a} is solved by
\begin{equation}\label{eom2aa}
\beta=\left(\frac{r}{r_{\infty}}\right)^p,
\end{equation}
where we have chosen a positive signature for $\beta$ without loss
of generality. According to equation \eqref{eom2e} we can write down
the relation $r^2\gamma_{,r}/\beta=-Q$ with a constant electric
charge $Q$, as demonstrated by \eqref{charge}. This relation implies
\begin{equation}\label{eom2ee}
\gamma=-\frac{Q}{r_{\infty}}\ln\left(\frac{r}{r_{\infty}}\right)
\end{equation}
for $p=1$ or
\begin{equation}\label{eom2eee}
\gamma=\frac{Q}{(1-p)r}\left(\frac{r}{r_{\infty}}\right)^p
\end{equation}
for $p\neq1$ up to an eliminable constant.

We will study the case with $f_{,T}=-1$ to some extent in subsection
\ref{subsect-open}. The case with $f_{,T}\neq-1$ will be
exhaustively investigated, for $p=1$ in subsection
\ref{subsect-conS}, and for $p\neq1$ in subsections \ref{subsect-RN}
($T_{,r}\neq0$) and \ref{subsect-Sch} ($T=\const$).

\subsection{$f_{,T}=-1$}\label{subsect-open}
In this case, the equations of motion are further simplified,
\begin{eqnarray}\label{eom21}
\nonumber 1+f_{,T}&=&0,\\
\nonumber f_{,TT}T_{,r}&=&0,\\
T+f&=&0.
\end{eqnarray}
If $T_{,r}\neq0$, equations \eqref{eom21} have only one solution
$f=-T$, resulting in an unphysical null Lagrangian. So we turn to
consider $T=\const$.

The expression of $T$ is given by \eqref{Ts}. It is difficult to
exhaust all of the solutions to $T_{,r}=0$, but we found a
particular solution
\begin{eqnarray}\label{sol2a}
\nonumber \beta&=&\frac{r}{r_{\infty}},\\
\nonumber \gamma&=&0,\\
\nonumber 1+f_{,T}&=&0,\\
T+f&=&0,
\end{eqnarray}
leaving $\alpha$ unconstrained. The corresponding metric is
\begin{equation}\label{metric2a}
ds^2=(1-\alpha)d\tilde{t}^2-\frac{r_{\infty}^2}{(1-\alpha)r^2}dr^2-r_{\infty}^2d\Omega^2.
\end{equation}
This metric generalizes the
$\mathbf{R}^1\times\mathbf{R}^1\times\mathbf{S}^2$ solution
\eqref{metric1a} of constant radius, but the constraints on
Lagrangian are the same. For this solution, the torsion scalar
$T=2/r_{\infty}^2$ and the Ricci scalar
$R=-(r^2\alpha_{,rr}+r\alpha_{,r}+2)/r_{\infty}^2$.

Another particular solution to \eqref{eom21} is
\begin{eqnarray}\label{sol2b}
\nonumber \alpha&=&\frac{\Lambda r^2}{3}+\frac{2GM}{r},\\
\nonumber \beta&=&1,\\
\nonumber \gamma&=&0,\\
\nonumber 1+f_{,T}&=&0,\\
T+f&=&0
\end{eqnarray}
with $T=-2\Lambda$. Here $\Lambda$ may be thought as the
``cosmological constant'' though this model does not include the
Einstein's gravity theory. Later in subsection \ref{subsect-Sch} we
will see such a solution actually exist for more general models with
any value of $f_{,T}$, which naturally incorporate the Einstein's
theory. Please refer to subsection \ref{subsect-Sch} for physical
interpretations of this solution.

\subsection{$f_{,T}\neq-1$, $p=1$}\label{subsect-conS}
In the previous subsection, we have studied some solutions with
$f_{,T}=-1$. From now on, we will consider the case of
$f_{,T}\neq-1$. In this case, one special example is $p=1$. Now the
solution of $\beta$ is given by \eqref{eom2aa} with $p=1$, while the
solution of $\gamma$ is \eqref{eom2ee}. Then one may directly check
that $T=2/r_{\infty}^2$ using equation \eqref{Ts}. Subsequently,
most equations of motion are automatically satisfied except for
\eqref{eom2c} and \eqref{eom2d}, which demand
\begin{equation}\label{sol2c1}
T+f=\frac{8\pi GQ^2}{r_{\infty}^4}
\end{equation}
and
\begin{equation}\label{sol2c2}
\alpha=\left[1-\frac{8\pi GQ^2}{r_{\infty}^2(1+f_{,T})}\right]\ln^2\left(\frac{r}{r_{\infty}}\right)+\ln\left(\frac{r}{r_{\infty}}\right)^q+C
\end{equation}
respectively. In the above solution, $q$ and $C$ are constants, and
we have taken into account the fact that $f_{,T}$ is independent of
$r$. It is easy to show that the corresponding geometry contains an
$\mathbf{S}^2$ sphere of constant radius $r_{\infty}$. More
interestingly, solution \eqref{sol1c} and constraints
\eqref{eom1bbb} can be regenerated from this solution by requiring
$\alpha=0$.

In what follows, restricting to the case $f_{,T}\neq-1$, we will
explore the possibility that $p\neq1$. We will put our results in
two subsections, according to whether the torsion scalar is a
constant.

\subsection{$f_{,T}\neq-1$, $p\neq1$, $T_{,r}\neq0$}\label{subsect-RN}
In this case $T$ is not a constant, so the solution to equation
\eqref{eom2b} is $T+f=(T-2\Lambda)/\lambda$. In order to give the
correct value of Newtonian constant, $\lambda=1$ should be imposed
on physically viable models. The viable model is simply the
Einstein's gravity theory with a cosmological constant $\Lambda$.
Although the model is not new, in this subsection we will briefly
cover it for completeness.

Since $f(T)$ is a constant, from equations \eqref{eom2c} and
\eqref{eom2d} we obtain
\begin{eqnarray}\label{eom2cd}
\nonumber T&=&-2\Lambda+\frac{4p}{r^2}\left(\frac{r}{r_{\infty}}\right)^{2p}-\frac{8\pi GQ^2}{r^4}\left(\frac{r}{r_{\infty}}\right)^{4p},\\
\alpha&=&\frac{p(p-2)}{(p-1)^2}+\frac{Cr^2}{(p-1)^2}\left(\frac{r_{\infty}}{r}\right)^{2p}+\frac{2GM}{(p-1)^2r}\left(\frac{r}{r_{\infty}}\right)^p-\frac{4\pi GQ^2}{(p-1)^2r^2}\left(\frac{r}{r_{\infty}}\right)^{2p}.
\end{eqnarray}
Here $M$ and $C$ are constants of integration, while $G$ is the
Newtonian constant. Comparing the solution with \eqref{Ts}, we can
identify $C=\Lambda/3$.

With the redefinition $\tilde{r}=r_{\infty}^pr^{1-p}$, this solution
can be reformed as
\begin{eqnarray}\label{sol2d}
\nonumber 1-\alpha&=&\frac{1}{(p-1)^2}\left(1-\frac{\Lambda \tilde{r}^2}{3}-\frac{2GM}{\tilde{r}}+\frac{4\pi GQ^2}{\tilde{r}^2}\right),\\
\nonumber \beta^{-2}r^2&=&\tilde{r}^2,\\
\nonumber \beta^{-2}dr^2&=&\frac{1}{(p-1)^2}d\tilde{r}^2,\\
\nonumber \gamma&=&\frac{Q}{(1-p)\tilde{r}},\\
f&=&-2\Lambda.
\end{eqnarray}

After the time coordinate is rescaled as
$\tilde{t}\rightarrow(1-p)\tilde{t}$, solution \eqref{sol2d}
recovers the familiar Reissner-Nordstr\"om metric, which describes a
charged black hole in de Sitter ($\Lambda>0$) or anti-de Sitter
($\Lambda<0$) or flat ($\Lambda=0$) spacetime. Other than the
coordinate transformation in this way, we can also recover the metric
directly from \eqref{eom2aa}, \eqref{eom2eee} and \eqref{eom2cd} by
setting $p=0$. In this solution, the integration constants $M$ and
$Q$ are mass and electric charge of the black hole respectively,
while $\Lambda$ is the cosmological constant.

\subsection{$f_{,T}\neq-1$, $p\neq1$, $T_{,r}=0$}\label{subsect-Sch}
Since $p\neq1$, the expressions of $\beta$ and $\gamma$ are given by
equations \eqref{eom2aa} and \eqref{eom2eee}. We insert them into
equation \eqref{eom2c} and expand the resulted equation in powers of
$r$,
\begin{equation}
2Tf_{,T}-f+T-\frac{4p}{r^2}(1+f_{,T})\left(\frac{r}{r_{\infty}}\right)^{2p}+\frac{8\pi GQ^2}{r^4}\left(\frac{r}{r_{\infty}}\right)^{4p}=0.
\end{equation}
The first two terms on the left hand side are constants, because $f$
and $f_{,T}$ are constants when $T$ takes a constant value. Remember
also that $f_{,T}\neq-1$, $p\neq1$, so this equation dictates $p=0$,
$Q=0$ and
\begin{equation}\label{eom2cc}
2Tf_{,T}-f+T=0.
\end{equation}
Therefore, the consistent solution is restricted to
\begin{equation}\label{sol2e1}
\beta=1,~~~~\gamma=0.
\end{equation}

Let us denote $T=-2\Lambda$ as a constant. Then equation \eqref{Ts}
can be solved as a differential equation of $\alpha$ by
\begin{equation}\label{sol2e2}
\alpha=\frac{\Lambda r^2}{3}+\frac{2GM}{r}.
\end{equation}
One may check that equations \eqref{eom2b} and \eqref{eom2d} are
satisfied automatically.

Extrapolating \eqref{eom2cc}, \eqref{sol2e1} and \eqref{sol2e2} to
the limit $f_{,T}=-1$, we reobtain equations \eqref{sol2b}, thus
equation \eqref{eom2cc} holds more generally as an existence
condition of this solution. The solution can be interpreted as a
Schwarzschild black hole in de Sitter ($\Lambda>0$) or anti-de
Sitter ($\Lambda<0$) or flat ($\Lambda=0$) spacetime. The
integration constant $M$ is related to the black hole mass, while
$\Lambda$ is a constant determined by the Lagrangian. Although
$\Lambda$ does not necessarily arise from a constant term in the
Lagrangian, we will still call it cosmological constant for
simplicity. It is straightforward to check that the Ricci scalar
$R=-4\Lambda$.

We observe that torsion scalar $T$ and equation \eqref{eom2cc}
depend only on ``cosmological constant'' $\Lambda$, so the mass
parameter $M$ does not enter into Lagrangian parameters. This is
consistent with our physical interpretations of the integration
constants. Equation \eqref{eom2cc} puts a constraint on Lagrangian
of viable models that admit such a solution. Since $T$ is a
constant, it is enough to require this equation to hold numerically,
not analytically. In other words, we do not have to solve it as a
differential equation.

In subsection \ref{subsect-MinkRRS}, we have found out a solution
describing the Minkowski spacetime. In frame \eqref{tetrad}, the
Minkowski solution exists in $f(T)$ theories satisfying $f(0)=0$
numerically. Actually, as can be confirmed directly, this solution
and its constraint on Lagrangian can be obtained as a limit $M=0$,
$\Lambda=0$ of the Schwarzschild solution in this subsection.

\section{Reconstruction of Lagrangian}\label{sect-lag}
The results in sections \ref{sect-regular} and \ref{sect-horizon}
look scattering. In the current section let us review them briefly
by collecting the existence conditions of some solutions. In other
words, we will review the conditions $f(T)$ should meet if certain
solutions exist in frame \eqref{tetrad}. As will be shown, some of
the results can be unified in the same form, although they were
derived under different assumptions. As a convention, when we say an
existence condition should hold ``analytically'', we mean the
equations should stand for any value of $T$, solved as a
differential equation analytically. Otherwise, the word
``numerically'' is used to mean that the equations are only to hold
for a specified value of $T$. We should stress that all of the
existence conditions here apply only to the frame \eqref{tetrad}.
But we do not claim anything about other frames. Particularly, when
the conditions are violated, these solutions may appear in a
different frame.

\subsection{Reissner-Nordstr\"om-(anti-)de Sitter Solution}\label{subsect-cond-RN}
We have investigated this solution in subsection \ref{subsect-RN}.
Using the notations of \eqref{tetrad}, this solution is
\begin{eqnarray}\label{sol3a}
\nonumber \alpha&=&\frac{p(p-2)}{(p-1)^2}+\frac{\Lambda r^2}{3(p-1)^2}\left(\frac{r_{\infty}}{r}\right)^{2p}+\frac{2GM}{(p-1)^2r}\left(\frac{r}{r_{\infty}}\right)^p-\frac{4\pi GQ^2}{(p-1)^2r^2}\left(\frac{r}{r_{\infty}}\right)^{2p},\\
\nonumber \beta&=&\left(\frac{r}{r_{\infty}}\right)^p,\\
\gamma&=&\frac{Q}{(1-p)r}\left(\frac{r}{r_{\infty}}\right)^p
\end{eqnarray}
with $p\neq1$. Solution \eqref{sol3a} exists in frame \eqref{tetrad}
if
\begin{equation}\label{cond3a}
f=-2\Lambda
\end{equation}
analytically, which is equivalent to the Einstein's theory with a
cosmological constant $\Lambda$. We achieved at this condition in
subsection \ref{subsect-RN} under the assumption $f_{,T}\neq-1$.
Nevertheless, by combining the fact that $T_{,r}\neq0$ for this
solution and the discussions at the beginning of subsection
\ref{subsect-open}, we can see the condition is robust even without
any assumption on $f_{,T}$.

\subsection{Schwarzschild-(anti-)de Sitter Solution}\label{subsect-cond-Sch}
The solution is
\begin{eqnarray}\label{sol3b}
\nonumber \alpha&=&\frac{\Lambda r^2}{3}+\frac{2GM}{r},\\
\nonumber \beta&=&1,\\
\gamma&=&0
\end{eqnarray}
in frame \eqref{tetrad}. The existence condition of solution
\eqref{sol3b} is
\begin{equation}\label{cond3b}
2Tf_{,T}-f+T=0
\end{equation}
numerically with a constant $T=-2\Lambda$. We can also consider
\eqref{cond3b} as the existence condition of (anti-)de Sitter or
Minkowski solution in that frame, depending on the value of
$\Lambda$. This solution has been studied in subsection
\ref{subsect-Sch}, which stands on a general ground. The special
case with $f_{,T}=-1$ was explored in subsection \ref{subsect-open},
while the limit of $M=0$, $\Lambda=0$ was investigated in subsection
\ref{subsect-MinkRRS}.

\subsection{Charged Solution with a Sphere of Constant Radius}\label{subsect-cond-cconS}
This solution is
\begin{eqnarray}\label{sol3c}
\nonumber \alpha&=&\left[1-\frac{8\pi GQ^2}{r_{\infty}^2(1+f_{,T})}\right]\ln^2\left(\frac{r}{r_{\infty}}\right)+\ln\left(\frac{r}{r_{\infty}}\right)^q+C,\\
\nonumber \beta&=&\frac{r}{r_{\infty}},\\
\gamma&=&-\frac{Q}{r_{\infty}}\ln\left(\frac{r}{r_{\infty}}\right)
\end{eqnarray}
in frame \eqref{tetrad}. Its existence condition is
\begin{equation}\label{cond3c1}
T+f=\frac{8\pi GQ^2}{r_{\infty}^4}
\end{equation}
numerically, where $T=2/r_{\infty}^2$. The above solution was
obtained in subsection \ref{subsect-conS}, but the result can be
extrapolated to the special case $\alpha=0$, namely the
$\mathbf{R}^1\times\mathbf{R}^1\times\mathbf{S}^2$ solution in
subsection \ref{subsect-RRS}. In that limit, there is an additional
existence condition
\begin{equation}\label{cond3c2}
1+f_{,T}=\frac{8\pi GQ^2}{r_{\infty}^2}.
\end{equation}
This condition can be successfully recovered from the requirement
$\alpha=0$. Further taking limit $Q/r_{\infty}=0$, we find the neutral
$\mathbf{R}^1\times\mathbf{R}^1\times\mathbf{S}^2$ solution
discussed in subsection \ref{subsect-MinkRRS}. The neutral solution
can be extended in another way, as to be elucidated in the coming
subsection.

\subsection{Neutral Solution with a Sphere of Constant Radius}\label{subsect-cond-nconS}
In frame \eqref{tetrad}, the solution is
\begin{eqnarray}\label{sol3d}
\nonumber \beta&=&\frac{r}{r_{\infty}},\\
\gamma&=&0,
\end{eqnarray}
leaving $\alpha$ unconstrained. This solution exists in frame
\eqref{tetrad} if the Lagrangian satisfies conditions
\begin{eqnarray}\label{cond3d}
\nonumber 1+f_{,T}&=&0,\\
T+f&=&0
\end{eqnarray}
numerically at $T=2/r_{\infty}^2$. We have investigated this
solution in subsections \ref{subsect-MinkRRS} ($\alpha=0$) and
\ref{subsect-open} ($\alpha\neq0$). Both solutions \eqref{sol3c} and
\eqref{sol3d} have a product space $\mathbf{S}^2$ of constant radius
$r_{\infty}$.

\subsection{Solution with a Sphere of Variable Radius}\label{subsect-cond-varS}
As has been discussed in subsection \ref{subsect-RRS}, the solution
\begin{eqnarray}\label{sol3e}
\nonumber \alpha&=&0,\\
\nonumber \beta&=&\frac{r^2}{r_{\infty}\sqrt{r^2+r_0^2}},\\
\gamma&=&\frac{Q}{r_{\infty}}\arcsinh\left(\frac{r_{\infty}}{r_0}\right)-\frac{Q}{r_{\infty}}\arcsinh\left(\frac{r}{r_0}\right).
\end{eqnarray}
has a product space $\mathbf{S}^2$ of variable radius
\eqref{radius}. The existence condition of this solution in frame
\eqref{tetrad} can be cast in the form
\begin{equation}\label{cond3e1}
T+f=\frac{8\pi GQ^2u(u+4)}{r_{\infty}^4(u+1)^2},
\end{equation}
which is expected to hold analytically. In condition
\eqref{cond3e1}, function $u(T)$ is implicitly given by the real
root of cubic equation
\begin{equation}\label{cond3e2}
T=\frac{2u(u+2)^2}{r_{\infty}^2(u+1)^3}.
\end{equation}
Note the second equation of \eqref{eom1bbbb} can be derived  from
the first one and equation \eqref{cond3e2}.

\subsection{Example of Lagrangian Reconstruction}\label{subsect-cond-app}
The existence conditions summarized above are useful. On the one
hand, given the Lagrangian of $f(T)$ model, we can quickly judge the
existence of these solutions in frame \eqref{tetrad}. On the other
hand, we can use them to reconstruct the Lagrangian admitting
certain solutions. Interestingly, some of the existence conditions
do not contradict with each other, so the corresponding solutions
could coexist in the same Lagrangian.

As a simplified example, let us construct a model that admits a de
Sitter solution, an anti-de Sitter solution and a Minkowski solution
at the same time. According to the existence condition in subsection
\ref{subsect-cond-Sch}, this can be realized by designing a function
$f(T)$ obeying
\begin{equation}
2Tf_{,T}-f+T\propto T(T+2\Lambda_1)(T+2\Lambda_2).
\end{equation}
Assuming the polynomial form of $f(T)$, we build a model
\begin{equation}\label{model3b}
f=\tilde{G}^2\left[T^3+\frac{10}{3}(\Lambda_1+\Lambda_2)T^2+20\Lambda_1\Lambda_2T\right]-T,
\end{equation}
where $\tilde{G}$ is a constant with the dimension of $\mass^{-2}$.
When $\Lambda_1\Lambda_2<0$, this model has the required multiple
solutions indeed.

\section{Discussion}\label{sect-concl}
Based on a special frame \eqref{tetrad}, we found some static
solutions with spherical symmetry in $f(T)$ gravity theories, where
the Maxwell term was taken into consideration. The solutions and
their existence conditions were summarized in section
\ref{sect-lag}. But there are several problems unsolved.

First, the tetrads \eqref{tetrc1} for FLRW spacetime are widely used
in the literature on $f(T)$ cosmology. The frame \eqref{tetrad} we
employed is the close cousin of \eqref{tetrc1}. However, as we have
shown in section \ref{sect-frame}, the reconstruction of Lagrangian
is very sensitive to the choice of frame, because the torsion scalar
$T$ is not locally Lorentz invariant. This poses the pressing
problem: why are the frames like \eqref{tetrc1} so special and what
shall we interpret the local Lorentz transformations in general
$f(T)$ theories? As an up-to-date reference, \cite{Ferraro:2011us}
provides a nice answer to this problem for FRLW spacetime, which is
further illustrated in \cite{Ferraro:2011zb}.

Second, our ansatz \eqref{tetrad} of tetrads simplifies the
computation greatly, but it also limits the validity of our results.
This means we have only found some spherically symmetric static
solutions for certain special $f(T)$ models. It is still an open
problem to get such solutions for more general models in other
frames. That would involve six more variables related to the local
Lorentz transformation, leading to equations of motion hard to
solve.

Third, in subsection \ref{subsect-open} we only laid out some
particular solutions to $T_{,r}=0$. Investigation is still needed to
exhaust all of its solutions.

Fourth, since the existence of black solutions has been confirmed in
this paper, the thermodynamics of gravity is to be studied in $f(T)$
theories.

\acknowledgments{The author would like to thank Qinyan Tan for
encouragement and support.}

\end{document}